\documentstyle[12pt,aas2pp4,psfig]{article}

\newcommand{\Teff}{$T_{\rm{eff}}$}

\newcommand{\Myr}{M$_{\odot}$~yr$^{-1}$}
\newcommand{\Ms}{M$_{\odot}$}
\newcommand{\kms}{km~s$^{-1}$}

\newcommand{\Zs}{Z$_{\odot}$}

\newcommand{\Mlow}{$M_{\rm{low}}$}
\newcommand{\Mup}{$M_{\rm{up}}$}
\tighten
 
\received{}
\accepted{Accepted for publication in ApJ}
\journalid{}{}
\articleid{}{}

\begin{document}

\title{Evolutionary synthesis modeling of 
red supergiant features in the near--infrared}
 
\author{Livia Origlia}
\affil{Osservatorio Astronomico di Bologna, 
        Via Zamboni 33, I--40126 Bologna, Italy\\
   e--mail: origlia@astbo3.bo.astro.it\\
     and\\
Space Telescope Science Institute, 3700 San Martin Drive, Baltimore,
       MD 21218}
 
\author{Jeffrey D. Goldader, Claus Leitherer, and Daniel Schaerer}
\affil{Space Telescope Science Institute, 3700 San Martin Drive, Baltimore,
       MD 21218\\
       e--mail: goldader@stsci.edu, leitherer@stsci.edu, schaerer@stsci.edu}
 
\and
 
\author{Ernesto Oliva}
\affil{Osservatorio Astrofisico di Arcetri, 
        Largo E. Fermi 5, I--50125 Firenze, Italy\\
   e--mail: oliva@arcetri.astro.it}
 
\begin{abstract}
 
We present evolutionary synthesis models applied to near--infrared spectral 
features observed in the spectra of young Magellanic Cloud 
clusters and starburst galaxies. 
The temporal evolution of the first and second overtones of CO at 2.29 \micron\ 
(2--0 bandhead) and 1.62 \micron\ (6--3 bandhead) 
and of the $(U-B)$, $(B-V)$ and $(J-K)$ colors
are investigated.  
 
We find that the current evolutionary tracks of massive stars with 
sub--solar chemical composition in the 
red supergiant phase are not reliable
for any synthesis of the temporal evolution of infrared stellar features. 

The high sensitivity of the selected infrared features 
to the atmospheric parameters of cool stars allows us to place  
constraints on the 
temperature and the fraction of time 
spent in the red part of the Hertzsprung--Russell diagram 
by massive stars during their core--helium burning phase. 

We derive a set of empirically calibrated spectrophotometric models by 
adjusting the red supergiant parameters such that the properties of the 
observed templates are reproduced. 
 
\end{abstract}
 
 
\medskip
	{\centerline\sl Accepted for publication in ApJ}

\section{Introduction}
 
Since Tinsley's (1972) pioneering work, various  
evolutionary synthesis models of stellar populations in galaxies 
have been developed, using improved theoretical and empirical spectral 
libraries for better comparisons between predicted and observed 
spectrophotometric properties of galaxies. 
A comprehensive compilation of the available models 
was published by Leitherer et al. (1996). 
 
Galaxies with dominant young stellar populations ($\la$100~Myr) are called
`starbursts' (Searle, Sargent, \& Bagnuolo 1973). An overview of the
properties of the stellar content of starburst galaxies can be found in
Leitherer (1996). Moorwood (1996) reviewed starburst galaxies with particular
emphasis on infrared (IR) aspects. 
 
Substantial theoretical and observational efforts have been made to 
calibrate suitable ultraviolet (UV) and optical diagnostics for studying 
the stellar content of starburst galaxies (Leitherer et al. 1996).
Massive stars have only few
and comparatively weak features in the optical spectral region.
Furthermore, most absorption lines are blended with nebular emission lines if
the stars are embedded in ionized gas. Notable exceptions are the strong
hot--star wind features in the UV (Leitherer, Robert, \& Heckman
1995) and the
Wolf--Rayet feature at 4650/4686~\AA, which has been used to define a subset
of starbursts called Wolf--Rayet galaxies (Conti 1991). Wolf-Rayet stars are
massive stars at the end point of their evolution (Abbott \& Conti 1987).
Attempts to utilize Wolf--Rayet features for constraining the stellar 
population have been made (e.g., Meynet 1995; Schaerer 1996; 
Schaerer \& Vacca 1998) 
although uncertainties in their modeling remain.
 
Only recently has such a systematic analysis been extended
to the near--IR range where red supergiants (RSGs) dominate. 
This spectral range is of particular
interest to the modeling of starburst spectra. 
RSGs are less evolved than Wolf--Rayet stars, and therefore 
their evolutionary status is thought to be better understood. 
The main features in the IR spectra of integrated stellar systems
are absorption lines due to neutral metals and molecules (e.g.,
Kleinmann \& Hall 1986; Origlia et al. 1993, hereafter OMO93)
tracing red stellar populations, and the hydrogen
recombination lines of the Paschen and Brackett series, which mainly
provide information on the number of ionizing photons.
Once the first RSGs have
formed, the near--IR spectra of starbursts show their characteristic
absorption features. Due to the fast technological improvements in IR array 
performances, near--IR spectra of stellar clusters and galaxies showing RSG
features are becoming available in increasing numbers (e.g., 
Lan\c{c}on \& Rocca--Volmerange 1992; Goldader et al. 1995, 1997; 
Oliva et al. 1995; Smith et al. 1996; Oliva \& Origlia 1998, 
hereafter OO98).
 
When dealing with integrated populations, one of the most 
critical issues in properly classifying their stellar content is the
disentanglement of age and metallicity effects (cf. e.g., the  
review by Renzini 1986 and, more recently, Worthey 1994). 
OMO93 discussed the behavior
of the near--IR CO features at $\lambda$1.62 and $\lambda$2.29 
in individual red stars with varying stellar parameters 
such as effective temperature, gravity, metallicity, and microturbulent 
velocity. 
The results of OMO93, combined with theoretical models for the dependence of 
these
parameters on stellar age and mass, offer the opportunity to make 
predictions for the strength of RSG spectral features 
during the times when massive RSGs dominate in starbursts, 
i.e. from about 8 to 30~Myr.
 
In this paper, we couple the new near--IR spectral diagnostics of OMO93 
to evolutionary synthesis models based on those of 
Leitherer \& Heckman (1995, hereafter LH95 and Leitherer et al. 1998, 
hereafter L98).
The technique
is outlined in \S~2. Results of the numerical simulations and the
confrontation with observations are in 
\S~3. Empirical modifications to the RSG models which lead to near--IR 
features in better agreement with observations are discussed in \S~4.  
Our conclusions are presented in \S~5.  
 
\section{Synthesis technique}
 
All models in this paper were constructed using an updated evolutionary 
synthesis code based on that of LH95. The new code is described in detail
by L98.  Briefly, the code now 
incorporates the most recent stellar evolutionary tracks from the Geneva
group at $Z$=0.04, 0.020, 0.008, 0.004, and 0.001 
(Schaller et al. 1992; Schaerer et al. 1993).
It has also been updated to
allow the use of isochrone synthesis.  Finally, the corrected 
grid of atmospheres compiled by Lejeune, Cuisinier \& Buser (1997) is used.
This atmosphere grid reproduces the observed effective temperature 
{\it versus} color relation at all metallicities.

As with the LH95 code, the new version allows 
the star formation rate to be one of two cases.
First, stars could form in an {\em instantaneous burst} from 
interstellar gas of mass $M_{\rm{tot}}$ with 
no subsequent star formation. 
Alternatively, the models allow for {\em continuous star formation,} 
at a constant level determined by the star formation rate in \Myr. 
The stellar population is
distributed along the zero--age main sequence following an intial mass 
function (IMF) which we parameterize as
\begin{equation}\phi(M) = \frac{dN}{dM} =  C M^{-\alpha}
\label{eq:imf}
\end{equation}
between the upper and lower cutoff masses, \Mup\ and \Mlow, respectively.
The normalization constant $C$ is determined by the
total stellar mass. $\alpha = 2.35$ corresponds to the classical Salpeter
(1955) value.   

Each star is tracked in the Hertzsprung--Russell diagram (HRD) from
the zero--age main sequence until its death. Effective temperatures \Teff,
surface gravities $\log g$, and surface abundances are assigned during each
time step and appropriate model atmospheres are chosen. 

[Z/Fe]=0 at all metallicities{\footnote{Throughout this paper we
define metallicity as the mass fraction of all elements heavier than helium.
The sun has Z=0.020 in this definition.}}
are adopted in both the model atmospheres and in the evolutionary models.
The nebular continuum 
(bound--free, free--free,
and two--photon) is computed and added to the stellar continuum. 
Thus, dilution of stellar spectral features 
by the nebular continuum is accounted for.
  
For the present study we enhanced the code by implementing the 
theoretical equivalent widths of CO $\lambda$1.62 and $\lambda$2.29
from OMO93. Their models predict these indices 
as a function of \Teff, $\log g$, microturbulent velocity $\xi$,  
and metallicities between 1/100 and 3 times solar.
These indices are successful in reproducing the corresponding 
observed spectra of both single giant and supergiant stars (OMO93), and
integrated stellar clusters with known parameters in both our Galaxy 
(Origlia et al. 1997) and the Magellanic Clouds (OO98). 
The equivalent widths were obtained at each time step from a 
multi--dimensional interpolation in the spectral grids.
 
At temperatures less than 3000~K we set the indices
to the value at the lowest tabulated temperature, with all other parameters
held constant; at temperatures higher than 4500~K, we set the indices to 
zero.
These assumptions are reasonable since at temperatures higher than 4500~K 
the CO molecule is almost dissociated, 
hence the equivalent widths of the selected features become very small.
Similarly, the CO temperature dependence becomes negligible below 3600~K
(OMO93), and the equivalent widths of the selected lines do not change 
significantly at lower temperatures. 
 
A large grid of models was run exploring the following parameter space:
\begin{itemize}
\item Instantaneous burst and continuous star formation models.
\item Ages $<$ 25 Myr. 
\item Metallicities $Z=0.020$ (=\Zs), $Z=0.008$ (=40\% solar).
\item IMF with slopes $\alpha = 2.35$ and 3.3.
\item \Mup~=~100~\Ms; \Mlow~=~1~\Ms.
\item Microturbulent velocities of 3 and 5~\kms, typical for RSGs 
      (e.g., Tsuji et al. 1994; OO98, and references 
      therein).
\item Solar and non--solar [C/Fe], where C was depleted  
      by 0.5~dex. Carbon is 
      expected to be slightly depleted (e.g., Lambert et al. 1984) 
      due to internal mixing in post--main--sequence stages.  
\end{itemize}

The changes in microturbulent velocity and [C/Fe] 
abundances were kept internal to the routine in which the values of the
indices were computed, and have no effect on other values calculated
by the code (e.g., colors).
 
The near--IR properties of stellar populations dominated by RSGs are rather
insensitive to variations of \Mup\ between 30 and 100~\Ms\ (cf. LH95). This
results from the small number of RSGs evolving from stars with progenitor
masses above $\sim$30~\Ms.

\section{Results from standard RSG models}
 
In this section we highlight the key results of our simulations.  
We show models for the CO $\lambda$1.62 and $\lambda$2.29 
features and $(U-B)$, $(B-V)$ and $(J-K)$ colors and we 
compare the results to two sets of observational data: 
a sample of Large Magellanic Cloud (LMC) clusters and 
starburst galaxies.  

\subsection{Observational templates}

We selected the 5 young LMC clusters NGC1818, NGC1984, NGC1994, NGC2004 and 
NGC2011 with measured CO features from OO98.
3 clusters from the OO98 sample were excluded.
NGC1866 and NGC1987 are too old to harbor massive RSGs,
and NGC330 is in the SMC with a much lower metallicity.
Optical colors are from van den Bergh (1981), 
extinction and the $(J-K)$ color are from Persson et al. (1983).
The uncertainties of the CO features and color are 0.3--0.5 \AA\ and 
$\le$0.1 mag, respectively, as discussed in the cited literature.
The same measurement errors apply to the starburst sample further below.
A range of ages has been assigned to each cluster according to the 
estimates from different authors (Hodge 1983; 
Elson \& Fall 1985, 1988; 
Girardi et al. 1995; Cassatella et al. 1996).  

The metallicity of individual young LMC clusters is not accurately known
(cf. e.g. the compilation by 
Sagar \& Pandey 1989) but on average values well below solar are inferred. 
For the selected clusters several independent estimates 
based on optical and infrared spectra 
(cf. Richtler, Spite \& Spite 1989; Reitermann et al. 1990; 
Jasniewicz \& Th\'evenin 1994; OO98) indicate values
between one tenth and at most half solar. 

The selected sample of starburst galaxies with measured spectral features 
is the one observed by Oliva et al. (1995). 
The galaxies are NGC253,
NGC1614, NGC1808, NGC3256, NGC4945, NGC7552, and NGC7714.
We also added NGC1705 to this sample, a well studied galaxy 
with colors from  Lamb et al. (1985) and Quillen, 
Ramirez \& Frogel (1995).
For these galaxies we also used $(U-B)$ and $(B-V)$ colors 
from Veron--C\`etty (1984) and Hamuy \& Maza (1987)
and the $(J-K)$ color from Glass \& Moorwood (1985). 
The colors were corrected for extinction assuming an 
intrinsic $(J-H)$ in the range 0.6--0.7 (cf. e.g., Scoville et al. 1985) 
and the interstellar extinction law by Rieke \& Lebofsky (1985) for a 
homogeneous dust distribution. 
The de--reddened values are quite uncertain due to the amount of the 
correction and its aperture--size dependence, 
particularly in the case of the $(U-B)$ and $(B-V)$.
Our main arguments will therefore be based on the equivalent widths of the 
near--IR features, which are reddening independent.
The presence of RSGs in the central region of these galaxies 
(Oliva et al. 1995) indicates 
a burst age (as traced by RSGs themselves and by the Brackett lines) 
in the range between 10 and 100~Myr.
The inferred light to mass ratios seem also to suggest that 
there is no significant contamination by an underlying, older 
stellar population, which would dilute the observed features.
Hot dust, if present, could dilute the spectral features as well.
OMO93 used the ratio of CO$\lambda$1.62/Si$\lambda$1.59 and 
CO$\lambda$1.62/CO$\lambda$2.29 to quantify the degree of dust dilution.
Oliva et al. (1995) applied this method to the starburst sample discussed 
in this paper and found no significant dust dilution 
in the central few arcseconds.

The metal abundance of starbursts  is poorly known and the available
estimates vary by up to an order of magnitude, depending on 
the element used to trace metallicity, the 
galaxy sampled region,
the observational technique and the spectral synthesis modeling.
As an example, for NGC253 values ranging between less than one tenth up to 
about solar have been proposed (cf. e.g. Webster \& Smith 1983; Carral et al. 
1994; Zaritsky, Kennicutt \& Huchra 1994; Arimoto et al 1997; Ptak 
et al. 1997; OO98).
Nevertheless, in our sample there at least two objects 
whose metallicity is better constrained 
in the range 0.3--0.5 solar, namely
NGC1705 (Storchi--Bergmann, Calzetti \& Kinney 1994) and NGC7714 
(Gon\-z\'a\-lez\---\-Del\-gado et al. 1995; Garc\'{\i}a--Vargas et al. 1997; OO98).
 
The star--formation histories in the LMC clusters are most likely close to 
the idealized case of an instantaneous burst, as suggested by their 
small sizes of a few pc. This makes them the simplest stellar systems for 
comparisons with models. 
Starburst galaxies are more complex, with typical size scales of 
$\sim$kpc. By averaging over such large regions, most of the time 
information is lost, and the star formation history can be  
approximated by a steady--state model.
The relatively small color and equivalent width range observed for starbursts 
suggest a steady--state situation as well. As opposed to the LMC cluster 
sample, the starbursts were not selected by their ages. Therefore it would 
be difficult to understand why they fall into a very narrow parameter range, 
which, for a given metallicity, would indicate equal ages in instantaneous 
models.
For simplicity and as two possible extreme situations
in the plots we compared pure instantaneous bursts with stellar clusters
and continuous star--formation models with galaxies.

\subsection{Spectroscopic indices and colors}

The temporal variation of the equivalent widths of the selected 
CO $\lambda$1.62 and $\lambda$2.29 features 
predicted by our models for both an instantaneous burst and continuous 
star formation using a Salpeter IMF are plotted in Figs.~1 and 2 
for solar and 40\% solar metallicities, respectively.
The corresponding values measured 
in the selected LMC clusters and starburst galaxies are 
plotted for comparison, as well.

The $(U-B)$, $(B-V)$ and $(J-K)$ colors of the 
$Z=0.020$ and $Z=0.008$ models and the corresponding, de--reddened  values 
measured in the LMC clusters and starburst galaxies are plotted 
in Fig.~3.
The relative large spread in color shown by the LMC clusters 
(a few tenths of magnitude) can be ascribed mainly to 
age and/or metallicity effects and some uncertainty in the reddening 
correction. 
In the case of starbursts the main uncertainty is the 
adopted reddening correction, as already mentioned above.      

>From the comparison of models and observational templates 
we find that the
observed CO features and $(J-K)$ colors 
can only be reproduced by models with metallicity 
very close to solar. Theoretical curves at 40\% solar metallicities
systematically predict too weak CO features and blue $(J-K)$ colors
for all objects
but  NGC1818 and NGC1984 which, however, are known
to have metallicities $\le$20\% solar.

The discrepancy between models and 
observations becomes 
even more severe when the latter are compared with 
continuous star formation models, which significantly underestimate 
(by up to a factor of two) the equivalent widths of the observed CO 
features and the $(J-K)$ color. 

Both the 40\% and solar metallicity models can reasonably fit the 
observed $(B-V)$ colors, while 
the behavior of the $(U-B)$ color is more complex,  
being redder than observed 
in the case of an instantaneous burst and bluer than observed in the 
case of continuous star formation.
 
\subsection{Why does the standard RSG model fail?}

By comparing theoretical predictions and observations of both 
LMC clusters and starburst galaxies with sub--solar metallicities, 
we have demonstrated that the near--IR features cannot be reproduced 
with models at appropriate metallicities.  

Different assumptions for the IMF do not change the conclusion, 
as one can see 
in Fig.~4 where we compare models with both a 
Salpeter and a Miller-Scalo IMF (approximated as a power--law with index 
3.30).
The results are very similar at both $Z=0.020$ and $Z=0.008$.
The negligible dependence on the IMF simply follows from the fact that RSGs 
provide both the line flux and the adjacent continuum. Therefore 
the IR indices sample a relative small stellar mass interval.
This is very different from, e.g., stellar--wind lines in the UV 
or the Wolf--Rayet feature at $\lambda$4686 \AA\ 
(cf. Leitherer et al. 1995; Schaerer \& Vacca 1998).

Different assumptions for the model of atmospheres 
do not change the result, either.
When we built the OMO93 code we performed tests to investigate  
how the choice of the adopted model of atmosphere
for given stellar parameters influences the computed
equivalent widths.
No significant dependence on the adopted model atmospheres was found.
The main reason is simply the fact that the main source of opacity 
in late type stars is the H$^-$ ion.
We also performed test calculations with our L98 code, which has 
the capability to model cool stars with Kurucz or Lejeune models, 
or with blackbodies.
All three choices uniformly give the same result: too weak 
RSG features.
As we will demonstrate below, the dominant effect is the failure of 
the evolutionary models which predict incorrect RSG parameters.
The failure is so severe that 
the details of the model atmospheres become negligible.

To understand why the model spectra differ from those of real
stellar clusters and galaxies, we should explore how RSGs affect the 
integrated light.
Before the first RSGs form, the near--IR spectrum will be dominated by
emission from ionized gas.  The UV spectrum is dominated by the
photospheric emission from O--B stars.  Once these stars begin to turn
off the main sequence, they become massive RSGs and dominate the 
near--IR spectrum in the range of ages between 8--30 Myr 
(involving stars of mass $\sim$25--10 M$_{\sun}$).
Actually the evolution of intermediate mass RSGs  
($\sim$5--15 M$_{\sun}$) is more complex since these 
stars also have blue excursions towards higher temperatures 
which may last a large fraction ($\ge$50\%) of the total 
core--helium burning phase, particularly at low metallicities.

During the RSG phase the integrated colors of the evolving 
stellar population should become red and the absorption 
features typical of cool stars deep, but how much red and 
deep depends on their average temperature and blue--loop duration, 
which in turn depends on many factors (metallicity, mass loss rate etc.)

In order to better identify the possible cause of the
mismatch between predicted and observed IR features, we
computed near--IR indices and colors of {\em individual stars} predicted
by the evolutionary models instead of considering an entire population. The
mass range of the relevant contributors to the IR indices and colors
at ages $\le$30 Myr is roughly 25 to 8~\Ms.
Fig.~5 shows the predicted CO indices of a 
25, 20, 15, 12, 10
and 8~\Ms\ star for solar and 40\% solar metallicity.

The \Zs\ models (cf. Fig.~5, left panels) 
for 25, 20 and 15~\Ms\ reach the RSG stage at
$\simeq$6.5, 8.5 and 12~Myr, respectively, when the equivalent
widths start to peak. About 1~Myr later the RSGs explode as supernovae,
and the CO feature is no longer observed. The duration of the RSG phase
increases with decreasing main--sequence mass as 
indicated by the broader
equivalent width peak at 15~\Ms\ as compared to that at 25 and 20~\Ms.
A qualitative change of this behavior occurs for stars with masses 
smaller than 12~\Ms. 
The 8--12~\Ms\ models at \Zs\ have two peaks,
separated by a more than 1~Myr long gap of zero equivalent width. The gap is
caused by a blue loop in the evolutionary models, when temperatures above
10,000~K are reached, too high to produce any CO
feature. 
Since RSGs have a blue excursion lasting about 
50\% of the entire RSG lifetime, no CO features are predicted during about 
50\% of the RSG lifetime for masses below about 15~\Ms.

Next we consider stars with 0.4~\Zs\ (cf. Fig.~5, right panels).
The resulting models produce lower equivalent widths than
at solar metallicity. 
Three main reasons can explain such a behavior:
at lower $Z$
the effective temperature
of RSGs is higher, blue loops start at larger masses ($\sim$15~\Ms), and 
they last longer (about 20\% more). 
The offset in time between the
indices of stars having the same mass but different $Z$ is due to slower
evolution at lower $Z$ (Maeder 1990).

In an instantaneous burst, integration over all mass ranges will lead to the
prediction that the presence or absence of blue loops significantly affects
the line equivalent widths and the $(J-K)$ color of a RSG population, 
while it does not severly contaminate the optical colors.
A clear example of this prediction is the discontinuity at about 14~Myr 
in the instantaneous burst models of the infrared features, where the 
equivalent widths of the absorption lines drop and the $(J-K)$ color turns 
blue for a short period of time (see Figs.~1 and 2). 
This discontinuity is also present in the solar models but it is less 
obvious since blue loops are more pronounced at low metallicity.

Cases for continuous star formation are affected to a much smaller
degree: the main contributor to the line equivalent width has a main--sequence
mass of 15~\Ms\ and above, and such stars do not have blue loops.

As a result of the warm temperatures and very pronounced 
blue loops at sub--solar $Z$ in the Geneva models,
the equivalent widths of the absorption features and the $(J-K)$ color 
cannot approach the values observed in the clusters.
Mayya (1997) already reached similar conclusions for the $(J-K)$ color. 
OO98, using different models based on Padova evolutionary tracks 
(Bertelli et al. 1994), also found that at low $Z$ the models predict 
too warm temperatures for the RSGs.

Stars on blue loops, which are still cooler 
than the more massive blue supergiants in an earlier evolutionary stage, 
significantly contribute to the $(U-B)$ color (mainly to the B luminosity) 
in an instantaneous burst. This makes the $(U-B)$ redder. 
The continuous star formation models predict much bluer $(U-B)$ than the 
instantaneous ones since the blue--loop stars are less dominant.
Moreover, the presence of the blue loops may also account for 
the apparent reddening of the $(U-B)$ color when metallicity decreases since 
the blue loops become progressively more pronounced.
  
While Meynet (1993) and Mayya (1997) suggested that 
perhaps changing the mass--loss rate 
could provide a solution, we have taken a more general approach.  
In an attempt to
provide guidance to the stellar evolution groups, we now derive
empirical limits on the RSG temperatures and the fraction of the
core--helium burning lifetimes the massive stars spent as RSGs (vs. 
time spent in blue loops, or as blue supergiants).

\section{Results from the modified models}
 
\subsection{Theoretical background}
 
There are a number of plausible reasons why the theoretical 
predictions in the near--IR do not match the observed RSG populations.
 
The coolest temperatures predicted by evolutionary models for RSGs
of fairly high mass ($M > 10$~\Ms) are quite uncertain.
The major uncertainty stems from the treatment of the external convective
layers in these stars.
As discussed by Maeder (1987), the treatment of non--adiabatic convection
with the usual mixing length theory (MLT) leads to supersonic convective
velocities, which is in contradiction to assumptions of this theory.
To remedy this situation one includes (cf. Schaller et al. 1992) 
the contribution of the acoustic flux to the energy transport and the 
turbulent pressure in the hydrostatic equilibrium. Furthermore, a density 
scale height is adopted to prevent density inversions.
The difference between this treatment and the usual MLT is illustrated
by Maeder \& Meynet (1987, their Fig.~9) and by Chiosi et al. (1992). 
In view of these uncertainties, the predictions for both the effective
temperatures of RSGs and their dependence on metallicity 
have to regarded as considerably uncertain. 
This most likely explains the difficulties of evolutionary models
to correctly reproduce the radii and temperatures of massive RSGs.
Future progress will be required to reliably model the 
extreme conditions in these convective envelopes.
 
Another crucial difference is the amount of time spent as a RSG
during the core--helium burning phase.
At solar metallicity a supergiant star spends a large fraction of its 
core--helium burning phase in the red  
near the coolest position it will reach on the HRD.  
As metallicity
decreases, more and more of the core--helium burning lifetime is spent at
significantly higher temperatures ($\ge 10^4$~K), where these
objects will presumably be classified as blue supergiants.
It is well known that the transition between blue and red supergiants is
particularly sensitive to different model ingredients (internal mixing, 
mass loss, opacities etc.; e.g., Maeder \& Meynet 1988; Ritossa 1996). 
As discussed in detail by Langer \& Maeder (1995) no 
current set of evolutionary models is able to reproduce correctly the 
{\em variation} of the observed ratio of blue to red supergiants with 
metallicity.

\subsection{Models with empirically adjusted RSG parameters}
 
Given the above situation, we explored a set of models where
the stellar effective temperature and the fraction of time spent by a
massive star in the red part of the HRD during the core--helium
burning phase are free parameters.
This should allow us to better quantify the role played by these parameters
in the overall spectral properties of the selected indices and
to provide feedback constraints on the evolutionary
tracks of massive stars.
We ran models at $Z=0.008$ fixing 
the fraction of time during the core--helium burning phase 
spent in the red (i.e. as RSG) at 10\%, 50\% and 98\% and 
the RSG temperature at 3500~K, 4000~K, and 4500~K, regardless 
its mass. 
The temperature and lifetime adjustments were performed as follows:
for all time steps on the evolutionary track after the step corresponding
to 90\%, 50\%, or 2\% of the total core--helium lifetime, 
the temperature was settled to 3500K, 4000K, or 4500K.

The main results are shown in Figs.~6--11             
for the CO features with both solar and sub--solar [C/Fe], 
$(U-B)$ and $(J-K)$ color models, for
both an instantaneous burst and continuous star formation.
The behavior of the CO feature at $\lambda$1.62 indicates that 
the observed values of young LMC clusters 
can be reproduced by 40\% solar metallicity models only in a 
limited range of 
parameters. Temperatures $<$4000~K and fraction of time spent in the red 
during the core--helium burning phase of at least $\sim$50\% are required. 
These can be regarded as upper and lower limits, respectively, 
when models with carbon depletion 
are considered. For the galaxies, similar temperatures and lifetimes are
required.  
It is also interesting to note that the 
discontinuity around 14 Myr progressively disappears 
when increasing the time spent in the red, also in better agreement 
with the modeling based on the Padova evolutionary tracks 
(cf. OO98). 
This is a clear indication that the discontinuity can be simply an artifact 
due to the blue loops and has nothing to do with the actual temporal evolution 
of the integrated stellar population.

The CO at $\lambda$2.29 is a less sensitive thermometer 
than the CO at $\lambda$1.62 but its behavior is fully consistent 
with the trends shown by the latter (cf. Figs.~8 and 9).
The analysis of the colors supports the temperature/RSG lifetime
conclusions from the spectroscopic features.  
Only in the same temperature/RSG lifetime regime
which can account for the indices, do we approach the $(J-K)=0.8-1$ needed
to account for the observations of clusters and galaxies (cf. Fig.~11).  

The $(U-B)$ color (cf. Fig.~10) becomes slightly bluer both 
increasing the lifetime and  
decreasing the temperature of RSGs. 
This behavior is the confirmation that blue loops    
being cooler than the bluest supergiants but significantly warmer than the 
RSGs have the net effect to 
make the $(U-B)$ redder and the $(J-K)$ colors bluer.
Going to lower temperatures and longer lifetimes in the RSG phase 
cause the models to have only very blue
and very red stellar populations dominating the luminosity 
in the UV to blue and in the infrared domain, respectively, in better 
agreement with the observed features.
 
Fractions of time spent in the red during the core--helium burning phase 
larger than 50\% imply a ratio of blue to red supergiants less than 1.
Direct estimates of this ratio from observed color--magnitude diagrams 
(CMDs) are very rare and severly affected by statistical effects. 
Nevertheless in the few published CMDs of young LMC clusters 
there is evidence that this ratio is truly $\le$1 even at relative low 
metallicities (e.g., Chiosi et al. 1995 for NGC330 and Balona \& 
Jerzykiewicz 1993 for NGC2004 and NGC2100). 
More efforts should be made to obtain a larger sample of 
reliable CMDs of young LMC clusters to provide the ratio of blue to red 
supergiants with much higher statistical significance.  
 
As shown by Mayya (1997) the observed Ca~II triplet, $(J-K)$ color and 
photometric CO index of a sample of young LMC clusters are better reproduced 
by models which enhance the mass loss 
by a factor of two with respect to the standard RSG model. 
However, the photometric CO index is still less than observed,    
and the UV--optical colors become redder than the observed ones. 
This indicates that mass loss cannot be the single cause for the 
discrepancy and other crucial parameters most likely play a 
major role, as discussed in Sect.~4.1.
Unless more details in the evolution of RSGs become clearer, 
any diagnostics which involve temporal evolution of the 
IR features 
is still too premature. For example the 
predicted decreases of the Ca~II triplet equivalent width and 
the CO photometric index in the age range between 12 and 100 Myr 
is most probably an artifact of too-pronounced blue loops in the 
Geneva evolutionary tracks.

\section{Conclusions}    
 
The key results of this study are:
\begin{itemize}
\item
Evolutionary tracks of RSGs can reasonably account for the 
observed optical features but fail to reproduce the observed 
$(U-B)$ color and the IR spectra of stellar clusters and starburst 
galaxies at {\em sub--solar metallicity}.
They do not produce sufficiently cool supergiants with long enough          
core--helium burning phase in the red part 
of the HRD.   
\item
The high sensitivity of the selected spectroscopic indices 
to the physical parameters of cool stars make the present simulations 
a powerful tool to place constraints on the temperature and fraction 
of time spent in the red: at 40\% solar metallicity, 
the former should be less than 4000~K, and the latter at least 50\%. 
\item         
Synthesis models for the burst evolution which use IR 
diagnostics of massive red stars with sub--solar metallicity 
are still unreliable due to deficiencies in stellar evolution models.
\item
Models with empirically adjusted RSG parameters allow 
a reasonable fit for both the observed UV to blue and IR features in young 
stellar clusters and starburst galaxies.
\item 
The full model set, with original RSG parameters at \Zs\ and  
with empirical adjustments at 40\% \Zs\, is available in electronic form 
at {\tt http://www.stsci.edu/

ftp/science/starburst}.  
\end{itemize}
 
\acknowledgments
 
Livia Origlia gratefully acknowledges financial support from the STScI Visitor 
Program. Funding for Jeff Goldader was provided by NASA grant GO--06672.01--95A from 
the Space Telescope Science Institute, operated by the Association of Universities 
for Reasearch in Astronomy, Inc., under NASA contract NAS5-2655. 
Claus Leitherer received financial support from the Visiting Professor 
Program of the University of Bologna. 
Salary support for Daniel Schaerer was provided by the STScI Director's 
Discretionary Research Fund.
The authors acknowledge the anonymous referee for the helpful comments and
suggestions.


\begin{figure}[htp]
\centerline{\psfig{figure=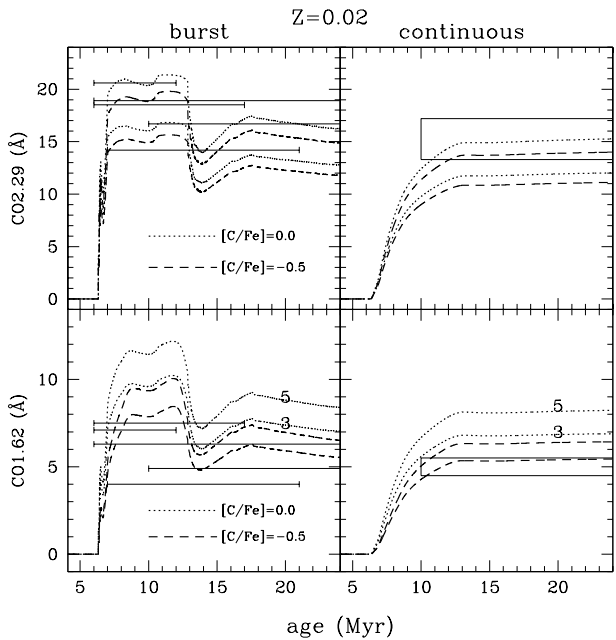,width=8.8cm}}
\label{fig1}
\caption{
The behavior of the CO spectroscopic features for instanteneous burst 
(left panels) and continous star formation models (right panels)
using a Salpeter IMF, solar metallicity, 
microturbulent velocity $\xi$ of 3 and 5 km/s,
solar [C/Fe] (dotted lines), one third solar [C/Fe] (dashed lines). 
The CO $\lambda$2.29 is related  to photometric CO index by the empirical 
relations (Kleinmann \& Hall 1986 and Origlia et al. 1997):              
$$ {\rm [CO]_{spec}} = -2.5 {\rm ~log~}
          \left [ 1 - {{\rm W_{\lambda}}(2.29)\over 53 {\rm \AA}} \right ];$$
$${\rm [CO]_{phot}} \simeq -0.6\times {\rm [CO]_{spec}} - 0.017.$$ 
For the instantaneous burst models we also show the values measured 
in a sample of young LMC clusters (cf. OO98), 
while in the continuous star formation models the rectangular boxes 
indicate the range of values ($\pm 1\sigma $) measured in a sample 
of starburst galaxies by Oliva et al. (1995).  
}
\end{figure}

\begin{figure}[htp]
\centerline{\psfig{figure=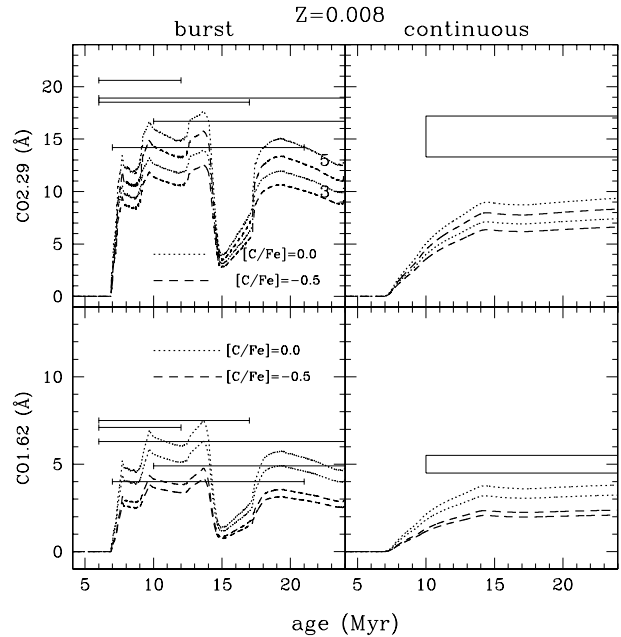,width=8.8cm}}
\label{fig2}
\caption{
As in Fig.~1, but for 40\% solar metallicity.
}
\end{figure}

\begin{figure}[htp]
\centerline{\psfig{figure=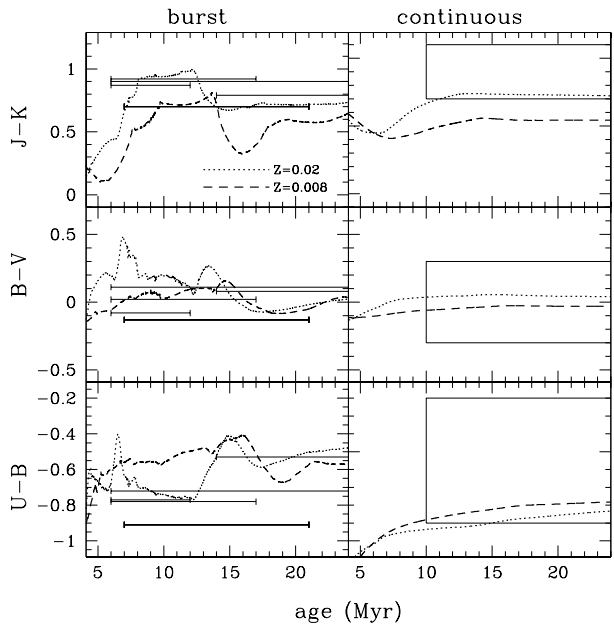,width=8.8cm}}
\label{fig3}
\caption{
Colors of the standard burst (left panels) and continuous star formation 
(right panels) models with $Z=0.020$ (dotted lines) 
and $Z=0.008$ (dashed lines). 
As in Figures 1 and 2, the de--reddened values of a sample of LMC 
clusters and starburst galaxies (boxes) are also shown for comparison.
The boxes indicate the range of values computed for the galaxies, 
assuming uncertainties in the adopted $A_V$ of $\pm$0.5 mag.
}
\end{figure}

\begin{figure}[htp]
\centerline{\psfig{figure=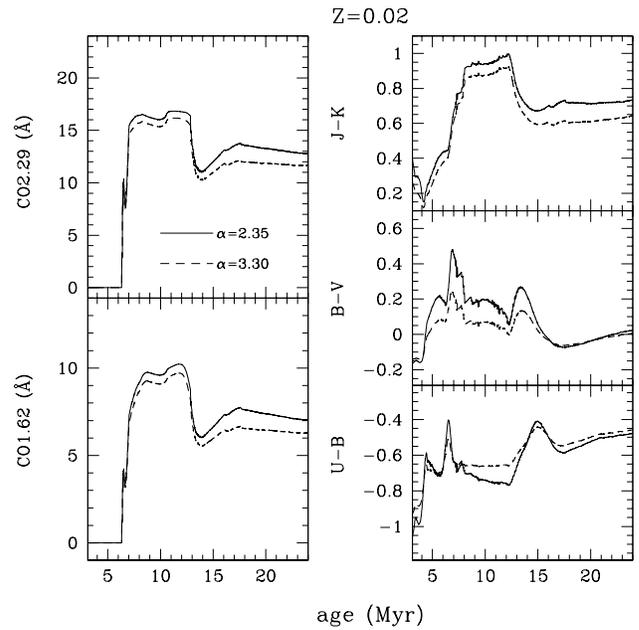,width=8.8cm}}
\label{fig4}
\caption{
Comparison between burst models with standard Salpeter (continuous lines) 
and Miller-Scalo (dashed lines, approximated by a power-law slope of 3.30) 
IMFs at $Z=0.02$. Differences in the
IMF have very little effect on the strengths of both the spectroscopic 
indices and colors.
}
\end{figure}

\begin{figure}[htp]
\centerline{\psfig{figure=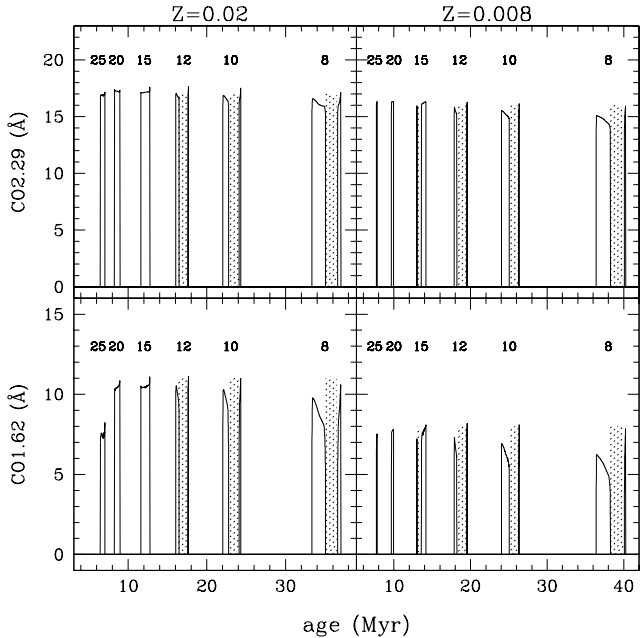,width=8.8cm}}
\label{fig5}
\caption{
Instantaneous burst models for the CO features, taking 
into account only the contribution of RSGs with single masses: 25, 20, 
15, 10 and 8 \Ms, respectively, at $Z=0.02$ (left panels) and $Z=0.008$ 
(right panels). The shaded regions correspond to the blue loops. 
}
\end{figure}

\begin{figure}[htp]
\centerline{\psfig{figure=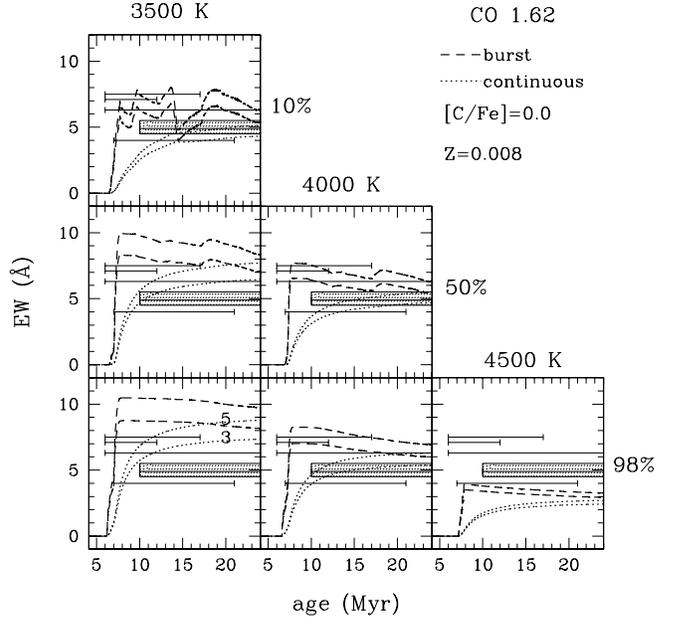,width=8.8cm}}
\caption{
\label{fig6}
Modified models for the CO $\lambda$1.62 index, 
where we have empirically adjusted the RSG temperature and 
lifetime during the core--helium burning phase.
The temperature is displayed at the top of the panels while the fraction 
of time spent as a RSG is displayed on the right side.  
In each panel, we show the instantaneous
burst (dotted lines) and continuous star formation (dashed lines) 
models at $Z=0.008$, solar 
[C/Fe] relative abundance and assume a Salpeter IMF.
The de--reddened values of young LMC clusters and starburst galaxies 
(shaded regtangles) are also plotted for comparison.
}
\end{figure}

\begin{figure}[htp]
\centerline{\psfig{figure=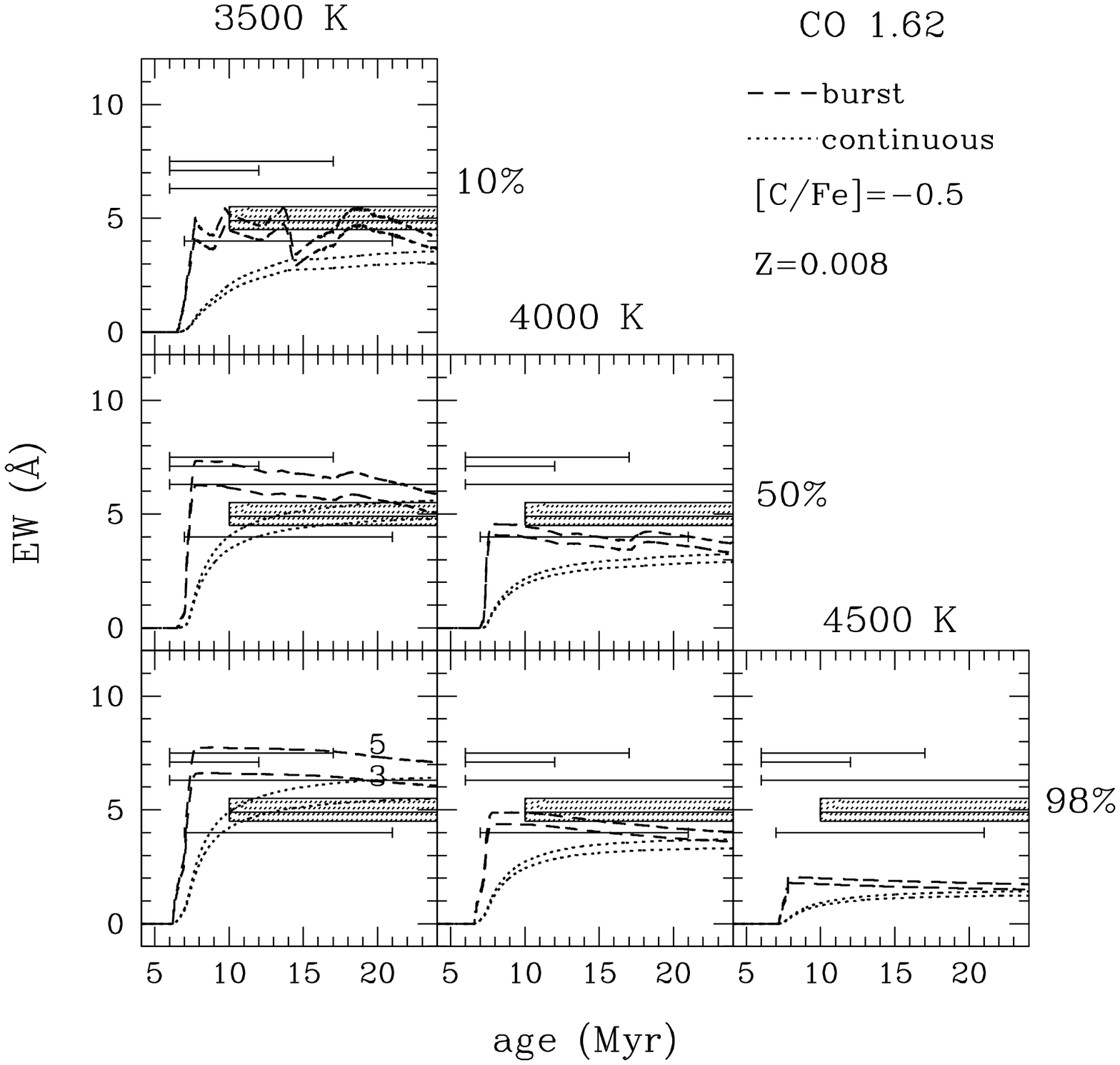,width=8.8cm}}
\label{fig7}
\caption{
As in Fig.~6, but for a [C/Fe]=--0.5.}
\end{figure}
 
\begin{figure}[htp]
\centerline{\psfig{figure=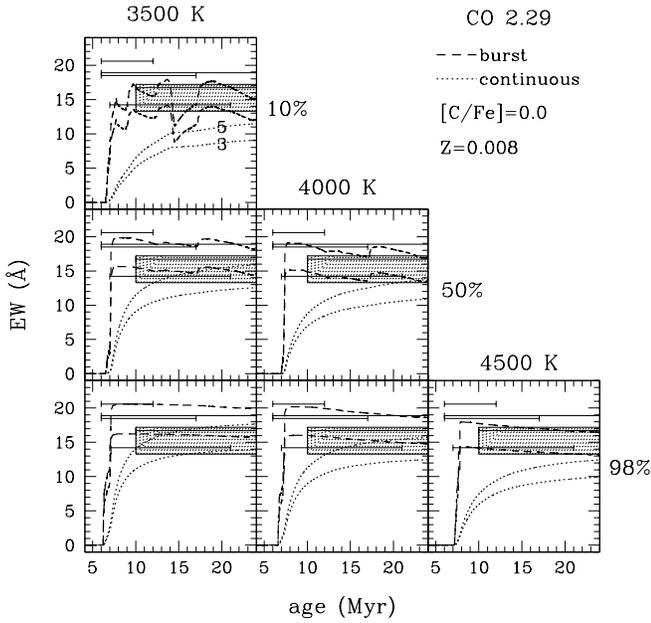,width=8.8cm}}
\label{fig8}
\caption{
As in Fig.~6, but for the CO $\lambda$2.29 index.}
\end{figure}

\begin{figure}[htp]
\centerline{\psfig{figure=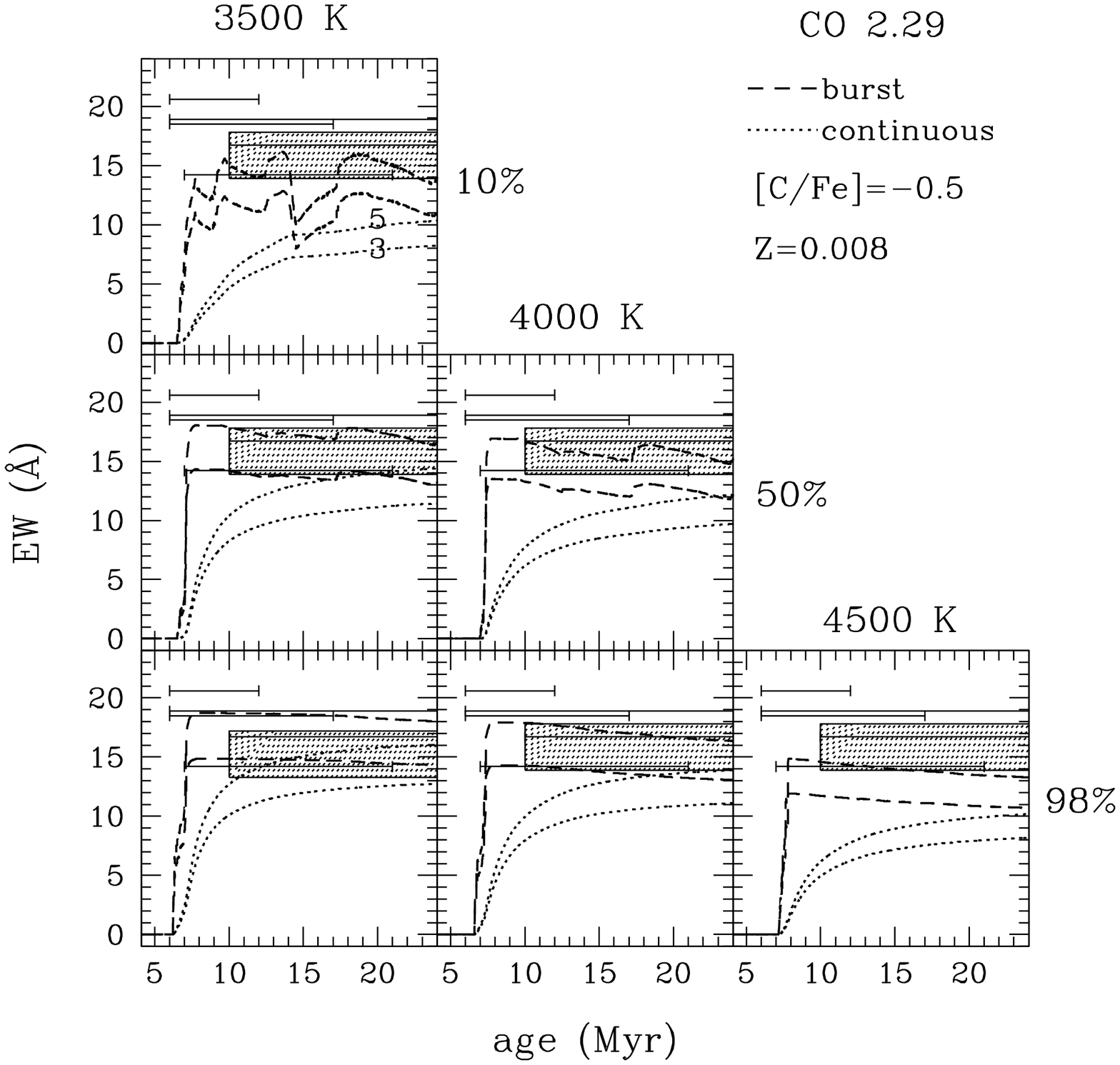,width=8.8cm}}
\label{fig9}
\caption{
As in Fig.~8, but for a [C/Fe]=--0.5.}
\end{figure}
 
\begin{figure}[htp]
\centerline{\psfig{figure=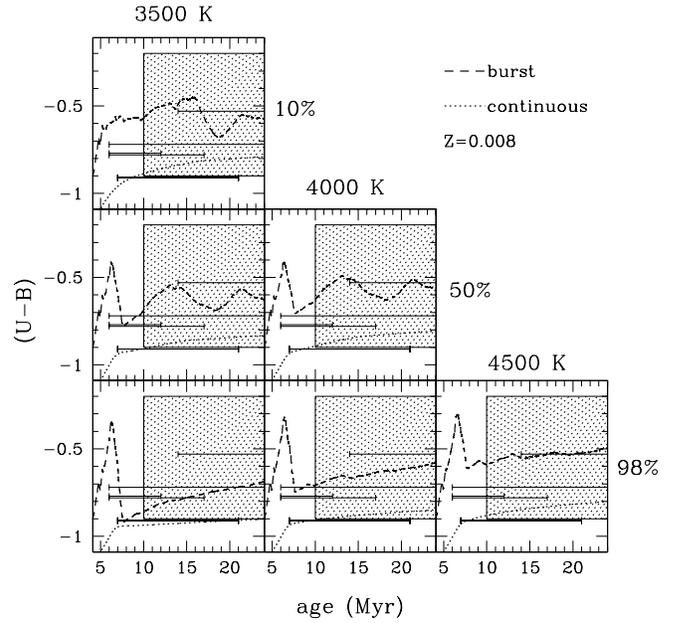,width=8.8cm}}
\label{fig10}
\caption{
As in Fig.~6 but for the $(U-B)$ colors.}
\end{figure}

\begin{figure}[htp]
\centerline{\psfig{figure=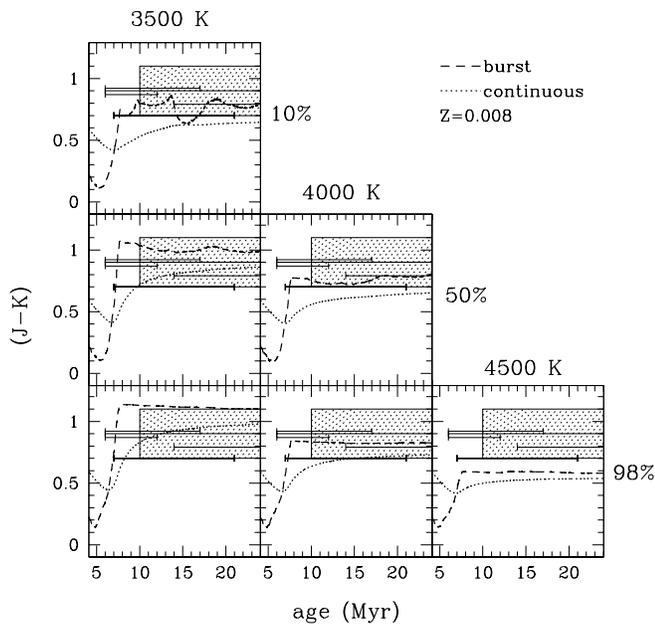,width=8.8cm}}
\label{fig11}
\caption{
As in Fig.~6, but for the $(J-K)$ colors.} 
\end{figure}


\end{document}